# Creating A Model HTTP Server Program

## Using java

Bala Dhandayuthapani Veerasamy

Department of Computing
Mekelle University
Mekelle, Ethiopia

*Abstract*— **HTTP Server is a computer programs that serves webpage content to clients. A webpage is a document or resource of information that is suitable for the World Wide Web and can be accessed through a web browser and displayed on a computer screen. This information is usually in HTML format, and may provide navigation to other webpage's via hypertext links. WebPages may be retrieved from a local computer or from a remote HTTP Server. WebPages are requested and served from HTTP Servers using Hypertext Transfer Protocol (HTTP). WebPages may consist of files of static or dynamic text stored within the HTTP Server's file system. Client-side scripting can make WebPages more responsive to user input once in the client browser. This paper encompasses the creation of HTTP server program using java language, which is basically supporting HTML and JavaScript.**

## I. INTRODUCTION

The Client-server [1] architecture is based on the principle where the 'client' program installed on the user's computer communicates with the 'server' program installed on the host computer to exchange information through the network. The client program is loaded on the PCs of users hooked to the Internet where as the server program is loaded on to the 'host' that may be located at a remote place. The concept of client-server computing has particular importance on the Internet because most of the programs are built using this design. The most important concepts and underlying mechanism that make the web works are Web Browser, Universal Resource Locators (URLs)[2], Hypertext Transfer Protocol (HTTP)[2], Hypertext Mark-up Language (HTML) [3] and Web Server [3].

Web browsers [3] are the applications that allow a user to view WebPages from a computer connected to the Internet. Web browser can read files created with WebPages and display them to the user. There are two important graphical browsers available for browsing WebPages [3], are Microsoft Internet Explorer [3] and Netscape Navigator. Most of the browser can be downloaded at without charge. The basic capabilities of a browser are to retrieve documents from the web, jump to links specified in the retrieved document, save, print the retrieved documents, find text on the document, and send information over the Internet. A web browser is a client program that uses the HTTP to make requests to the HTTP

Servers on behalf of the user. Web documents are written in a text formatting language called Hypertext Mark-up Language (HTML) [3]. The HTML is used to create hypertext documents that can be accessed on the web. Basically it is a set of 'mark-up' tags or codes inserted in a web file that tells the web browser how to display a web page for the user.

The Hypertext Transfer Protocol (HTTP) [2] is a set of rules for exchanging hypermedia documents on the World Wide Web. Hypermedia [3] simply combines hypertext and multimedia. Multimedia is any mixture of text, graphics, art, sound, animation and video with links and tools that let the person navigate, interact, and communicate with the computer. The web browser is an HTTP client, sending requests to server machines. When a user requests for a file through web browser by typing a Uniform Resource Locator then browser sends HTTP request that the destination server machine receives the request and, after any necessary processing, the requested file is returned to the client web browser.

The URL [3] is a compact string representation for a resource available on the Internet. URLs contain all of the information needed for the client to find and retrieve a HTML document such as protocol, domain name or IP address and webpage. Every HTTP Server has an IP address and usually a domain name, e.g. **www.mu.edu.et**. Server software runs exclusively on server machines, handling the storage and transmission of documents. In contrast, client software such as, Netscape, Internet Explorer, etc. runs on the end-user's computer accessing, translating and displaying documents.

A server is a computer system that is accessed by other computers and / or workstations at remote locations. A web server [3] is a software or program that process HTML documents for viewing by web browsers such as IIS, Apache HTTP Server [5] and WebLogic Server [6]. The server enables users on other sites to access document and it sends the document requested back to the requesting client. The client interprets and presents the document. The client is responsible for document presentation. The language that web clients and servers use to communicate with each other is called the HTTP. All web clients and servers must be able to communicate HTTP in order to send and receive hypermedia documents. For this reason, web servers are often called HTTP servers.

## II. HYPER TEXT MARKUP LANGUAGE

The term HTML is an acronym that stands for Hypertext Markup Language [3]. You can apply this markup language to your pages to display text, images, sound and movie files, and almost any other type of electronic information. You use the language to format documents and link them together, regardless of the type of computer with which the file was originally created.

HTML is written as plain text that any Web browser can read. The software does this by identifying specific elements of a document (such as heading, body, and footer), and then defining the way those elements should behave. These elements, called tags, are created by the World Wide Web Consortium (W3C). Most HTML tags come in pairs. You use the first tag in the pair (for example, <html>) to tell the computer to start applying the format. The second tag (for example, </html>) requires a slash in front of the tag name that tells the computer to stop applying the format. The first tag is usually referred to by the name within the bracket (for example, HTML). You can refer to the second tag as the end, or the close, tag (for example, end HTML).

HTML is a plain text file and needs a simple text editor to create the tags. However, it is important that all HTML documents have the extension .html or .htm which is three / four letter extension. Windows 'Notepad' may be used as an editor for writing the HTML files. Every HTML document should contain certain standard HTML tags. These tags describe the overall structure of a document, identify the document to browsers and provide simple information about the document. These structure tags do not affect its appearance and they are essential for tools that interpret HTML files. These structural elements are:

<HTML>

<HEAD>

<TITLE>Creating model HTTP Server</TITLE>

</HEAD>

<BODY>

. . . the document . . .

</BODY>

</HTML>

The <HTML> tag indicates that the content of the file is in the HTML language. All the text and commands in a document should go within the beginning and ending HTML tags. The <HEAD> tag specifies a limited amount of bibliographic data related to the document. It is the second item in the document. This element does not contain any text that displays in the browser except for the title that appears in the title bar of the browser. Each HTML document needs a title to describe the content of the document. The title is used by the browser to display it in its title bar. The <BODY> tag follows the HEAD tag. It contains all parts of the document to be displayed in the browser.

There are several number of tags avail for developing a webpage. Here few important tags are discussed. Headings are used to divide sections of text, like in any document. They are used to designate the logical hierarchy of the HTML document. There are currently six levels of headings defined. The number indicates heading levels (<H1> to <H6>). Each heading will have closing tags. When it displayed in a browser, will display differently. We can use paragraph tag <P> to indicate a paragraph. A browser ignores any indentations or blank lines in the source text. Without a <P> tag, the document becomes one large paragraph. The paragraph tag indicates a plain text paragraph. However, many browsers expect the opening paragraph tag to indicate the end of a paragraph. The horizontal rule tag, <HR>, has no closing tag and no text associated with it. The <HR> tag creates a horizontal line on the page. It is excellent for visually separating sections on your web page. It is often seen at the end of text on web pages and before the address information. For example see program 1.

Program 1. HTML program

<!—index.html -->

<HTML>

<HEAD>

    <TITLE>HTTP Server</TITLE>

</HEAD>

<BODY>

    <HR>

    <H1 align=center> Welcome to HTTP Server</H1>

    <H3 align=center> Using Java</H3>

    <HR>

    <H5 align=center> Developed by Bala Dhandayuthapani Veerasamy</H5>

</BODY>

</HTML>

The above program can be saved as index.html, it can be produced the result on local computer web browser as the following Fig.1.

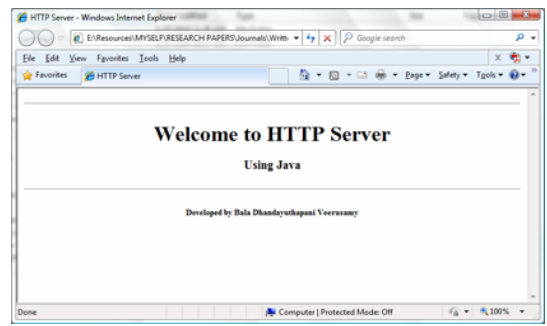

Figure 1. Output of the HTML

### III. USING JAVA NETWORKING CONCEPT

TCP and IP [2] together manage the flow of data, both in and out, over a network. TCP is a standard protocol with STD number 7. TCP is described by RFC 793 – Transmission Control Protocol. Its status is standard, and in practice, every TCP/IP implementation that is not used exclusively for routing will include TCP. TCP provides considerably more facilities for applications than UDP. Specifically, this includes error recovery, flow control, and reliability. TCP is a *connection-oriented* protocol, unlike UDP, which is *connectionless*. UDP is a standard protocol with STD number 6. UDP is described by RFC 768 – User Datagram Protocol. Its status is standard and almost every TCP/IP implementation intended for small data units transfer or those which can afford to lose a little amount of data will include UDP.

The Java networking package [7], also known as java.net, contains classes that allow you to perform a wide range of network communications. The networking package includes specific support for URLs, TCP sockets, IP addresses, and UDP sockets. The Java networking classes make it easy and straightforward to implement client/server Internet solutions in Java. The java networking package included web interface classes**,** raw network interface classes and extension classes**.** This study focuses on raw networking class such as Socket, ServerSocket, DatagramSocket, and InetAddress. These classes are providing access to plain, bare-bones networking facilities.

A Socket class is the Java representation of a TCP connection. When a Socket is created, a connection is opened to the specified destination. Stream objects can be obtained to send and receive data to the other end. Socket class constructors take two arguments: the name (or IP address) of the host to connect to, and the port number on that host to connect to. The host name can be given as either a String or as an InetAddress object. In either case, the port number is specified as an integer.

Socket( String host, int port )

Here, Socket constructor takes the hostname as IP address of the destination machine and port as the destination port to contact. The two most important methods in the Socket class are getInputStream() and getOutputStream(), which return stream objects that can be used to communicate through the socket. A close() method is provided to tell the underlying operating system to terminate the connection.

A ServerSocket class represents a listening TCP connection. Once an incoming connection is requested, the ServerSocket object returns a Socket object representing the connection. In normal use, another thread is spawned to handle the connection. The ServerSocket object is then free to listen for the next connection request. The constructors for this class take as an argument the local port number to listen to for connection requests and it also takes the maximum time to wait for a connection as a second argument.

ServerSocket( int port,int count )

ServerSocket takes the port number to listen for connections on and the amount of time to listen. The most important method in the ServerSocket class is accept(). This method blocks the calling thread until a connection is received. A Socket object is returned representing this new connection. The close() method tells the operating system to stop listening for requests on the socket.

### IV. A MODEL HTTP SERVER PROGRAM

There are several HTTP Servers are available to access the webpage such as Personal Web Server, Internet Information Server, Apache HTTP Server and etc. The program 2 created listening 8080 port to access WebPages on present working folder. Obviously present folder will act like www folder. The program 1 will have to store in the present folder, where we saving the following HTTPServer.java program. This HTTP Server program will support HTML and JavaScript, because both of them can be default understood by any web browser without having additional library.

Program 2. HTTP Server

```
//HttpServer.java
import java.net.*;
import java.io.*;
import java.util.*;

class HttpRequest
{
private Socket ClientConn;
public HttpRequest(Socket ClientConn) throws Exception
{
     this.ClientConn=ClientConn;
}
public void process() throws Exception
{
     DataInputStream din=new
     DataInputStream(ClientConn.getInputStream());
     OutputStream ot=ClientConn.getOutputStream();
     BufferedOutputStream out=new
     BufferedOutputStream(ot);

     String request=din.readLine().trim();
     StringTokenizer st=new StringTokenizer(request);
     String header=st.nextToken();

     if(header.equals("GET"))
     {
          String name=st.nextToken();
          int len=name.length();
          String fileName=name.substring(1,len);

          FileInputStream fin=null;
          boolean fileExist=true;

          if(fileName.equals(""))
          fileName="index.html";

          try
          {
```

```
fin=new FileInputStream(fileName);
}
catch(Exception ex) {      fileExist=false; }

String ServerLine="Simple HTTP Server";
String StatusLine=null;
String ContentTypeLine=null;
String ContentLengthLine=null;
String ContentBody=null;

if(fileExist)
{
StatusLine="HTTP/1.0 200 OK";
ContentTypeLine="Content-type:
   text/html";

ContentLengthLine="Content-Length: "+
   (new Integer(fin.available()).toString());

int temp=0;
byte[] buffer = new byte[1024] ;
int bytes = 0 ;
while ((bytes = fin.read(buffer)) != -1 )
{
out.write(buffer, 0, bytes);
for(int iCount=0;iCount<bytes;iCount++)
{
temp=buffer[iCount];
}
 }
fin.close();
}
else
{
StatusLine = "HTTP/1.0 200 OK";
ContentTypeLine="Content-type:
   text/html";

ContentBody = "<HTML>"
+ "<HEAD><TITLE>404 Not Found</TITLE> </HEAD>"
+ "<BODY><center><h1>404: The file "
+ fileName +" is not found" + "</h1></center> </BODY>
</HTML>" ;

out.write(ContentBody.getBytes());
}

out.close();
ClientConn.close();

}

}
}
class HttpServer
{
public static void main(String args[]) throws Exception
{
System.out.println("\n\n\t\tThe HTTP Server is running..");
System.out.println("\n\n\t\tStop server using  Ctrl + C");
```

```
ServerSocket soc=new ServerSocket(80);

while(true)
{
   Socket inSoc=soc.accept();
   HttpRequest request=new HttpRequest(inSoc);
   request.process();
      }
    }
}
```

## V. SETTING UP CONNECTIONS

When most people think of a firewall [8], they think of a device that resides on the network and controls the traffic that passes between network segments. However, firewalls can also be implemented on systems themselves, such as with Microsoft Internet Connection Firewall (ICF), in which case they are known as host-based firewalls. Fundamentally, both types of firewalls have the same objective: to provide a method of enforcing an access control policy. Indeed, at the simplest definition, firewalls are nothing more than access control policy enforcement points. Firewalls enable you to define an access control requirement and ensure that only traffic or data that meets that requirement can traverse the firewall or access the protected system. Firewalls need to be able to manage and control network traffic, authenticate access, act as an intermediary, protect resources, record and report on events. The first and most fundamental functionality that all firewalls must perform is to manage and control the network traffic that is allowed to access the protected network or host. Firewalls typically do so by inspecting the packets and monitoring the connections that are being made, and then filtering connections based on the packet-inspection results and connections that are observed. While executing HTTP server program first time, the firewall security alert will appear in that we should unblock. This is shown in the figure 2.

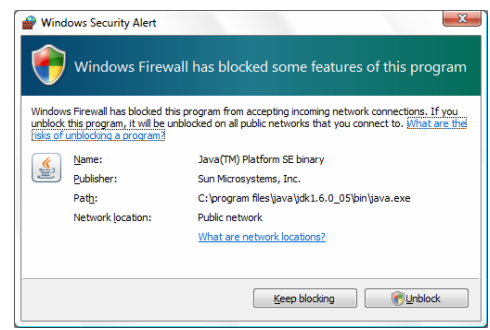

Figure 2. Firewall Settings

To compile the HTTP server program, use javac HttpServer.java and to execute the HTTP server program at the server computer, use java HttpServer. After executing the program one console window will appear, this is ready to share web document. We can press Ctrl+C to close the HTTP server. (See figure 3).

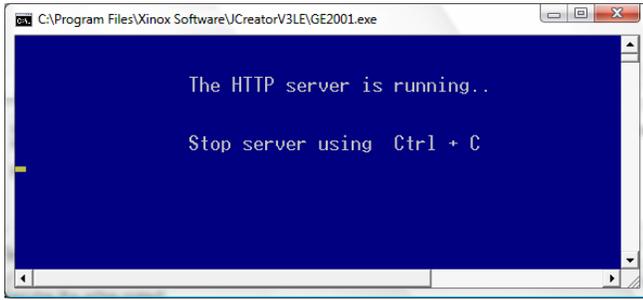

Figure 3. Running HTTP Server

## VI. RESULT AND DISCUSSION

As you know, this model HTTP Server is created using java. First of all we needed to install JDK 1.4 [9] or later version and JCreator 3.50 [10], before compiling and executing this program. This program complied and executed by JCreator 3.50 software, which is the tool enabling us to write program, compile and execute easily. HTTPServer.java has been tested only in window operating system under the Local Area Network in Mekelle University. Here I have used a HTTP server with the IP of http://10.128.40.145; hence we should access WebPages only through this IP. The following figure 4 is a sample result, which I got on network client. This server brought the index.html page by default. The index.html has been stored in the present working folder of HTTP server.

I hope it will also support other operating system well, because java is a platform independent language. Hyper Text Transfer Protocol views any webpage only by using port number 80. So this HTTP server created listening on port number 80. Hence it can capable to access HTML and JavaScript basically, which are defiantly understood by any web browsers. HTML provides creating static WebPages and JavaScript allowed to have client side scripting for any user input validations. This HTTP Server would not support for any server side scripting. In future, I do have a plan to develop own library to support for the server side scripting.

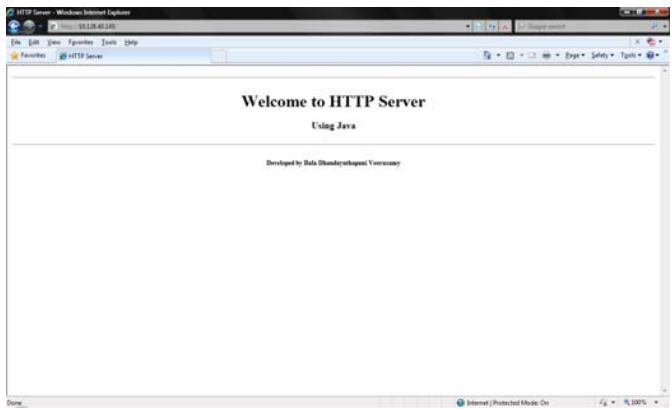

Figure 4. Trial Testing

## VII. CONCLUSION

HTTP Server is a computer programs, written by using java language. It just serves HTML WebPages with static content only to any client over any network. It also support for client-side scripting that can make WebPages more receptive to user input on the client browser; the client side scripting can mostly understood by any web browsers. This paper emphasizes how to create a new HTTP server using java language. It brought basic idea, wherein researchers may enhance further more developments to support for the already available server side scripting or may think on producing library to support innovative server side scripting in future.

## REFERENCES

[1] Steve Steinke, Network Tutorial, Fifth Edition, CMP Books, 2003

[2] Libor Dostálek, Alena Kabelová, Understanding TCP/IP,Packt Publishing, 2006.

[3] Deidre Hayes, Sams Teach Yourself HTML in 10 Minutes, Fourth Edition, Sams, 2006.

[4] Palmer W. Agnew, Anne S. Kellerman, Fundamentals of Multimedia, IGI Global, 2008.

[5] Apache HTTP Server, http://www.apache.org/

[6] BEA WebLogic Server, http://www.bea.com/

[7] Elliotte Rusty Harold, Java Network Programming, O' Reilly, 2000.

[8] Wes Noonan, Ido Dubrawsky, Firewall Fundamentals, Cisco Press, 2006

[9] Herbert Schildt, Java™ 2: The Complete Reference, Fifth Edition, McGraw-Hill, 2002.

[10] JCreator, http://www.jcreator.com/

AUTHORS PROFILE

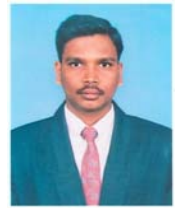

**Bala Dhandayuthapani Veerasamy** was born in Tamil Nadu, India in the year 1979. The author was awarded his first masters degree M.S in Information Technology from Bharathidasan University in 2002 and his second masters degree M.Tech in Information Technology from Allahabad Agricultural Institute of Deemed University in 2005. He has published more than fifteen peer reviewed technical papers on various international journals and conferences. He has managed as technical chairperson of an international conference. He has an active participation as a program committee member as well as an editorial review board member in international conferences. He is also a member of an editorial review board in international journals.

He has offered courses to Computer Science and Engineering, Information Systems and Technology, since 8 years in the academic field. His academic career started in reputed engineering colleges in India. At present, he is working as a Lecturer in the Department of Computing, College of Engineering, Mekelle University, Ethiopia. His teaching interest focuses on Parallel and Distributed Computing, Object Oriented Programming, Web Technologies and Multimedia Systems. His research interest includes Parallel and Distributed Computing, Multimedia and Wireless Computing. He has prepared teaching material for various courses that he has handled. At present, his textbook on "An Introduction to Parallel and Distributed Computing through java" is under review and is expected to be published shortly. He has the life membership of ISTE (Indian Society of Technical Education).